\documentclass{jpp}
\usepackage{graphicx, color}

\usepackage[utf8]{inputenc}
\usepackage[T1]{fontenc}
\usepackage{amsmath,comment}
\usepackage[capitalize]{cleveref}

\newcommand{\vect}[1]{\mbox{\boldmath $#1$}}

\newcommand{\etabar}{\bar{\eta}}
\newcommand{\sign}{\mathrm{sign}}
\newcommand{\sgn}{\mathrm{sgn}}

\newcommand{\changed}[1]{{#1}}
\newcommand{\revII}[1]{{#1}}

\newcommand{\DMerc}{D_{\mathrm{Merc}}}
\newcommand{\DShear}{D_{\mathrm{Shear}}}
\newcommand{\DWell}{D_{\mathrm{Well}}}
\newcommand{\DCurr}{D_{\mathrm{Curr}}}
\newcommand{\DGeod}{D_{\mathrm{Geod}}}

\shorttitle{Magnetic well and Mercier stability of stellarators near the magnetic axis}
\shortauthor{M Landreman and R Jorge}

\title{Magnetic well and Mercier stability of stellarators near the magnetic axis}

\author{Matt Landreman\aff{1}
  \corresp{\email{mattland@umd.edu}}
 Rogerio Jorge\aff{1}  }

\affiliation{\aff{1}Institute for Research in Electronics and Applied Physics,
University of Maryland,
College Park, MD 20742, USA
}

\begin{document}

\maketitle

\begin{abstract}
We have recently demonstrated that by expanding in small distance from the magnetic axis compared to the major radius, stellarator shapes with low neoclassical transport can be generated efficiently. To extend the utility of this new design approach, here we evaluate measures of magnetohydrodynamic interchange stability within the same expansion.
In particular, we evaluate magnetic well, Mercier's criterion, and resistive interchange stability near a magnetic axis of arbitrary shape.
In contrast to previous work on interchange stability near the magnetic axis, which used
an expansion of the flux coordinates, here we 
use the `inverse expansion' in which the flux coordinates are the independent variables.
Reduced expressions are presented for the magnetic well and stability criterion in the case of quasisymmetry.
The analytic results are shown to agree with calculations from the VMEC equilibrium code.
Finally, we show that near the axis, Glasser, Greene, \& Johnson's stability criterion for resistive modes approximately coincides with Mercier's ideal condition.
\end{abstract}


\section{Introduction}

The geometry of stellarators can be optimized to achieve properties such as low neoclassical transport and good magnetohydrodynamic (MHD) stability.
In the design of recent experiments like W7-X \citep{Beidler1990}, HSX \citep{HSX}, and NCSX \citep{Zarnstorff}, this optimization was done by wrapping a 3D MHD equilibrium code with a
standard numerical minimization algorithm.
\changed{This approach typically yields a single local minimum in the objective function without providing information about the possible existence of other local minima. Weights are chosen to combine multiple objectives into a single objective function, and one is left wondering whether a different choice of weights might lead to promising magnetic configurations in a different region of parameter space.
}
An older approach for relating stellarator geometry to physics properties
is to make an asymptotic expansion in large local aspect ratio \citep{Mercier1964,SolovevShafranov,MercierLuc,LortzNuhrenberg,GB1}. 
Such an expansion is accurate in the core of any stellarator, even those for which the aspect ratio of the boundary is low \citep{GarrenBoozerConsistency}.
We have recently argued \citep{PaperII,r2GarrenBoozer,Jorge2} that this asymptotic approach deserves further attention, as it complements numerical optimization. The asymptotic approach allows equilibria to be evaluated orders of magnitude faster, and it provides a practical way to generate new initial conditions for numerical optimization \changed{\citep{PaperII,r2GarrenBoozer,PaperIII}}.
In the present paper we extend our asymptotic approach, which has previously focused on neoclassical confinement, to MHD stability. Focusing on interchange modes, we show how stability can be computed directly from a solution of the reduced near-axis equations. These results enable more comprehensive design within the near-axis approximation.

A condition for stability of radially localized ideal-MHD interchange modes, known as `Mercier's criterion', was derived by \cite{MercierNFSupplement,Mercier1964} 
and \cite{GreeneJohnson1961,GreeneJohnson1962}. An important related quantity is the `magnetic well'.
In the absence of a pressure gradient, the magnetic well is $d^2 V / d \psi^2$, where $V(\psi)$ is the volume enclosed by a flux surface and $2\pi\psi$ is the toroidal flux. When the pressure is nonuniform, several generalized expressions for magnetic well can be defined \citep{Greene}. As shown by \cite{Mercier1964}, the magnetic well is the largest term in Mercier's criterion if one expands in large aspect ratio and makes a subsidiary expansion in $2 \mu_0 p / B^2 \ll 1$. (This ratio is not assumed to be small in our calculations here.)
Mercier stability and magnetic well are both commonly used in stellarator design \citep{HSX,ROSE}.
Mercier's criterion can be generalized to the case of nonzero plasma resistivity,
giving a stricter stability condition \citep{Glasser1975}.

Already from Mercier's original work on the ideal stability criterion, the limit of the criterion near the magnetic axis has been examined. However such previous work has generally employed the `direct' expansion, in which the function $\psi(\rho)$ is expanded, with $\rho$ the Euclidean distance from the axis \citep{Mercier1964,SolovevShafranov,Jorge1}. We instead follow the `inverse' expansion of \cite{GB1,GB2}, in which the flux coordinates are the independent variables, and the position vector is expanded as a function of these variables.
The direct and indirect expansions each have advantages and disadvantages. Here we focus on Garren \& Boozer's expansion because interchange stability has not previously been analyzed in this approach, and because it is convenient for obtaining configurations with omnigenity and high-order quasisymmetry \citep{PaperIII,r2GarrenBoozer}.

Following Mercier's original work
on near-axis stability, many researchers examined the stability criterion near the axis in axisymmetry 
\citep{Laval1971, KuppersTasso1972, LortzNuhrenberg1973, Mikhailovskii, Weimer}. These results for axisymmetry have been reviewed by \cite{Greene} and \cite{FreidbergMHDNew}.
In nonaxisymmetric geometry,
the magnetic well near the axis in vacuum was examined by \cite{Whiteman}. 
Later work on stellarator Mercier stability near the axis, \revII{reviewed by \cite{Pustovitov},} has mostly
examined special cases such as that of constant elongation \citep{ShafranovYurchenko}, circular cross-section \citep{MikhailovskiiAburdzhaniya},
or a planar axis \citep{Rizk}.

It should be acknowledged that Mercier stability may not be critical experimentally. The LHD, W7-AS, and TJ-II stellarator experiments all have operated in Mercier-unstable regimes, without obvious strong experimental signatures when stability boundaries are crossed \citep{Geiger,Watanabe,Weller2006,Aguilera}. These studies provide some indications that turbulence may increase in Mercier-unstable regimes. These observations also suggest that Mercier instabilities in stellarators may saturate at low amplitude. Nonetheless, one may still wish to include Mercier stability as a condition in new designs to minimize any possible turbulent transport associated with this class of instability. 

The remainder of this paper is organized as follows.
We begin the detailed calculations in the next section by defining variables and reviewing the asymptotic expansion. Then in section \ref{sec:well}
we compute several variants of magnetic well.
Mercier's criterion is evaluated in section \ref{sec:Mercier}. For each of these last two
sections, we present expressions for both a general
stellarator and a quasisymmetric one, as 
a number of simplifications occur in quasisymmetry.
Sections \ref{sec:well} and \ref{sec:Mercier} also
include demonstrations that our near-axis expressions
agree with finite-aspect-ratio calculations using the VMEC code \citep{VMEC1983} in the appropriate limit.
Then in section \ref{sec:resistive}, resistive interchange stability is examined, and it is shown
that the the criterion of \cite{Glasser1975}
coincides with Mercier's condition near the axis.
Previously published expressions for Mercier stability have often been derived assuming that quantities such as the toroidal flux and/or Jacobian are positive; in the appendix we show how these expressions generalize to allow other signs.

\section{Notation}
\label{sec:notation}

We will use the expansion developed by \cite{GB1} and the notation of \cite{r2GarrenBoozer}.
The notation and expansion are summarized here for convenience.
Let $\theta$ and $\varphi$ denote the Boozer poloidal and toroidal angles respectively, and let $2 \pi \psi$ be the toroidal flux. Then the magnetic field can be written
\begin{align}
\label{eq:BoozerCoords}
\vect{B} = &\nabla\psi \times\nabla\theta + \iota \nabla\varphi \times\nabla\psi, \\
 = &\beta \nabla\psi + I \nabla\theta + G \nabla\varphi, \nonumber
\end{align}
where $I$ and $G$ are constant on flux surfaces. In case
one wishes to consider quasi-helical symmetry, it is convenient to introduce a helical angle
$\vartheta = \theta - N \varphi$ where $N$ is a constant integer; $N$ can be set to zero if not considering quasi-helical symmetry. Defining $\iota_N = \iota - N$, then
\begin{align}
\vect{B} = &\nabla\psi \times\nabla\vartheta + \iota_N \nabla\varphi \times\nabla\psi,
\label{eq:straight_field_lines_h}
\\
 =& \beta \nabla\psi + I \nabla\vartheta + (G+NI) \nabla\varphi.
\label{eq:Boozer_h}
\end{align}
\changed{By using these expressions we are assuming that magnetic surfaces exist, which is not always the case for nonaxisymmetric fields. This assumption is reasonable for the stability calculations here because surfaces always exist in some neighborhood of the magnetic axis, and surface quality would be addressed separately by e.g. design of the $\iota$ profile.
}

At any location in the plasma we can express the position vector $\vect{r}$ as
\begin{align}
\label{eq:positionVector}
\vect{r}(r,\vartheta,\varphi) = \vect{r}_0(\varphi)
+X(r,\vartheta,\varphi) \vect{n}(\varphi)
+Y(r,\vartheta,\varphi) \vect{b}(\varphi)
+Z(r,\vartheta,\varphi) \vect{t}(\varphi).
\end{align}
Here $\vect{r}_0(\varphi)$ is the position vector along the magnetic axis, $r(\psi)$ is an effective minor radius defined by $2 \pi \psi = \pi r^2 \bar{B}$, and  $\bar{B}$ is a constant reference field strength of the same sign as $\psi$. The Frenet-Serret frame of the axis $(\vect{t},\vect{n},\vect{b})$ is a set of orthonormal vectors satisfying
$\vect{t}\times\vect{n}=\vect{b}$
and
\begin{align}
\frac{d\varphi}{d\ell}
\frac{d\vect{r}_0}{d\varphi} = \vect{t}, 
\hspace{0.3in}
\frac{d\varphi}{d\ell}
\frac{d\vect{t}}{d\varphi} = \kappa \vect{n}, 
\hspace{0.3in}
\frac{d\varphi}{d\ell}
\frac{d\vect{n}}{d\varphi} = -\kappa \vect{t} + \tau \vect{b}, 
\hspace{0.3in}
\frac{d\varphi}{d\ell}
\frac{d\vect{b}}{d\varphi} = -\tau \vect{n}.
\label{eq:Frenet}
\end{align}
Here $\ell$ is the arclength along the axis, $\kappa(\varphi)$ is the axis curvature, and $\tau(\varphi)$ is the axis torsion. (The opposite sign convention for torsion is used by Garren and Boozer.)

Let $\mathcal{R}$ denote the scale length of the axis, i.e. $\mathcal{R} \sim 1/\kappa \sim 1/\tau$, and let $\epsilon = r/\mathcal{R}$.
We now expand in $\epsilon \ll 1$.
The coefficients $X$, $Y$, and $Z$ are expanded in the following way:
\begin{align}
X(r,\vartheta,\varphi)
= r X_1(\vartheta,\varphi) + r^2 X_2(\vartheta,\varphi) + r^3 X_3(\vartheta,\varphi) + \ldots.
\label{eq:radial_expansion}
\end{align}
We expand $B$ and $\beta$ in the same way but with an $r^0$ term:
\begin{align}
\label{eq:radial_expansion_B}
B(r,\vartheta,\varphi)
= B_0(\varphi) + r B_1(\vartheta,\varphi) + r^2 B_2(\vartheta,\varphi)+ r^3 B_3(\vartheta,\varphi) + \ldots,
\end{align}
The radial profile functions 
$G(r)$, $I(r)$, $p(r)$, and $\iota_N(r)$
must be symmetric under $r \to -r$, so only even powers of $r$
are included in their expansions:
\begin{align}
p(r) = p_0 + r^2 p_2 + r^4 p_4 + \ldots.
\end{align}
The profile $I(r)$ is proportional to the toroidal current inside the surface $r$, so $I_0 = 0$.
The magnetic field must be smooth, so as discussed in appendix A of \cite{PaperI}, the expansion coefficients must have the form
\begin{align}
\label{eq:poloidal_expansions}
X_1(\vartheta,\varphi) = &X_{1s}(\varphi) \sin(\vartheta) + X_{1c}(\varphi) \cos(\vartheta), \\
X_2(\vartheta,\varphi) = &X_{20}(\varphi) + X_{2s}(\varphi) \sin(2\vartheta) + X_{2c}(\varphi) \cos(2\vartheta), \nonumber \\
X_3(\vartheta,\varphi) = &X_{3s3}(\varphi) \sin(3\vartheta) + X_{3s1}(\varphi) \sin(\vartheta) + X_{3c3}(\varphi) \cos(3\vartheta) + X_{3c1}(\varphi) \cos(\vartheta).
 \nonumber
\end{align}
The same form applies to the expansion coefficients of $Y$, $Z$, $B$, and $\beta$.

The vectors $\nabla r$, $\nabla\vartheta$, and $\nabla\varphi$ are obtained from derivatives of the position vector (\ref{eq:positionVector}) using the dual relations. The results are substituted into
the equations (\ref{eq:straight_field_lines_h}) $=$ (\ref{eq:Boozer_h}) and $\vect{J}\times\vect{B} = \nabla p$, where $\mu_0 \vect{J} = \nabla\times\vect{B}$. The quasisymmetry condition $B(r,\vartheta,\varphi)=B(r,\vartheta)$ can also be imposed if desired. The result is a set of equations at each order in $\epsilon \ll 1$. These equations are displayed in \cite{GB2} and appendix A of \cite{r2GarrenBoozer}.

For the calculations that follow, it is useful to 
introduce symbols for the signs of two quantities:
$s_G = \sgn(G) = \pm 1$, and $s_\psi = \sgn(\psi) = \sgn(\bar{B}) = \pm 1$. Each of these signs can be flipped individually by reversing
the signs of the poloidal or toroidal angle,
as discussed in the appendix.

In the case of quasisymmetry, $d B_0/d\varphi=0$ so it is convenient to choose the
reference field as $\bar{B} =s_\psi B_0$. Also the poloidal angle can be chosen so $B_{1s}=0$, and we will use Garren \& Boozer's symbol $\etabar=B_{1c}/B_0$. We can understand $\etabar$ as a measure of the 
field strength variation, $B = B_0[ 1 + r \etabar \cos\vartheta + O(\epsilon^2)]$.


\section{Magnetic well}
\label{sec:well}

In stellarator optimization, the magnetic well \citep{Greene,FreidbergMHDNew} is often
included as a fast-to-evaluate proxy for MHD stability \citep{HSX,ROSE}.
For completeness, here we will consider three expressions for magnetic well that appear in the literature: $V''=d^2 V/d\psi^2$,
\begin{align}
\hat{W} = \frac{V}{\left< B^2 \right>}\frac{d \left<B^2\right>}{d V},
\end{align}
and
\begin{align}
W = \frac{V}{\left< B^2 \right>}
\frac{d}{dV} \left< 2 \mu_0 p + B^2 \right>.
\label{eq:W}
\end{align}
Here $\left< \ldots \right>$ denotes a flux surface average,
and $V(\psi)$ is the volume enclosed by the surface $\psi$. Throughout this paper we consider  $V$ to be non-negative, hence 
\begin{align}
V(\psi) = s_\psi \int_0^{\psi} d\psi \int_0^{2\pi} d\vartheta \int_0^{2\pi} d\varphi |\sqrt{g}|,
\label{eq:volume}
\end{align}
with $\sqrt{g} = (G+\iota I)/B^2$. 
For any quantity $Q$, the flux surface average is
\begin{align}
\left< Q \right> = \frac{s_\psi |G+\iota I|}{V'} \int_0^{2\pi} d\vartheta \int_0^{2\pi}d\varphi \frac{Q}{B^2}
\end{align}
where $V'=dV/d\psi = s_\psi |G+\iota I| \int_0^{2\pi} d\vartheta \int_0^{2\pi}d\varphi \; B^{-2}$.
Using $\left< B^2 \right> = 4 \pi^2 s_\psi |G+\iota I|/V'$, the first two definitions of magnetic well are related by
\begin{align}
\hat{W} = -\frac{V V''}{(V')^2}
+ \frac{V}{V'} \frac{d \ln |G+\iota I|}{d\psi}.
\label{eq:What}
\end{align}
Negative $V''$ is favorable for stability,
as is positive $\hat{W}$ or positive $W$.

Note that in vacuum, $W=\hat{W}$ and the last term in (\ref{eq:What})
vanishes, leaving the ratio $\hat{W}/V''$ equal to 
the negative-definite quantity $-V/(V')^2$. Therefore the three measures
of magnetic well provide equivalent information
in the limit of small plasma pressure.


\subsection{First form of magnetic well}

The first quantity we consider is $V''$.
To evaluate this quantity, we begin with (\ref{eq:volume}).
We insert the near-axis expansions and apply $d^2/d\psi^2$, noting
$d/d\psi = (r \bar{B})^{-1} d/dr$, $\sign(\bar{B})=s_\psi$,
and eq (A50) of \cite{r2GarrenBoozer}. We thereby obtain
\begin{align}
V''= 2 \pi \left| \frac{ G_0}{\bar{B}}\right|
\int_0^{2\pi}d\varphi \frac{1}{B_0^4} \left[3\left(B_{1s}^2+B_{1c}^2\right) -4 B_0 B_{20}
-\frac{\mu_0 p_2 B_0^2}{\pi}\int_0^{2\pi} \frac{d\hat{\varphi}}{B_0(\hat{\varphi})^2}
\right]
+O(\epsilon^2).
\label{eq:well1_general}
\end{align}
Here, $B_{1s}$, $B_{1c}$, $B_{20}$, and $B_0$ are functions of $\varphi$, except where $B_0$ is evaluated at $\hat\varphi$ as noted.
This formula applies to any toroidal plasma, not only quasisymmetric ones.
Notice that the leading-order magnetic well depends on the $O(\epsilon^2)$ variation of $B$, so a $O(\epsilon^1)$ solution is not accurate enough to compute the well.

In the special case of quasisymmetry, $B_0$ becomes independent of $\varphi$, and we take $\bar{B} = s_\psi B_0$. Also $B_{1s}=0$, and $B_{1c} = \etabar B_0$.
While $B_{20}$ is independent of $\varphi$ in quasisymmetry,
it is convenient to relax this requirement for practical construction of quasisymmetric configurations, as described in \cite{r2GarrenBoozer}. The elimination of the requirement $d B_{20}/d\varphi=0$ makes it possible to obtain solutions for any shape of the magnetic axis. Therefore here we will not demand that $B_{20}$ be independent of $\varphi$. We thus obtain
the following expression for the magnetic well in quasisymmetry:
\begin{align}
\label{eq:wellQS}
V''= \frac{4 \pi^2  |G_0|}{B_0^3}
\left[ 3 \etabar^2 - \frac{4 \bar{B}_{20}}{B_0} - \frac{2 \mu_0 p_2}{B_0^2}\right] +O(\epsilon^2),
\end{align}
where $\bar{B}_{20} = (2\pi)^{-1} \int_0^{2\pi}d\varphi B_{20}$.


\subsection{Second form of magnetic well}

The alternative magnetic well quantity $\hat{W}$ in
(\ref{eq:What}) can be evaluated in the near-axis
expansion by substituting the series for $B$ into
\begin{align}
    \left<B^2\right> = 4\pi^2 \left(
    \int_0^{2\pi}d\vartheta \int_0^{2\pi}d\varphi \frac{1}{B^2}\right)^{-1},
\end{align}
keeping terms through $O(\epsilon^2)$. Then substituting the result into (\ref{eq:What}),
one obtains
\begin{align}
\hat{W} = 2 r^2 \left(\int_0^{2\pi}\frac{d\varphi}{B_0^2}\right)^{-1}
\int_0^{2\pi} \frac{d\varphi}{B_0^4} \left[-\frac{3}{4}(B_{1s}^2+B_{1c}^2)+  B_{20}B_0\right],
\label{eq:What_nearAxis}
\end{align}
for a general toroidal plasma.
In the special case of quasisymmetry, this expression reduces to
\begin{align}
\hat{W} = 2 r^2\left( -\frac{3}{4}\bar{\eta}^2 +  \frac{\bar{B}_{20}}{B_0}\right).
\label{eq:What_QS}
\end{align}


\subsection{Third form of magnetic well}

Evaluating $(2\mu_0 V/\left<B^2\right>) dp/dV$ near the axis and adding the result to (\ref{eq:What_nearAxis}) gives an expression for (\ref{eq:W}):
\begin{align}
W = 2 r^2 \left(\int_0^{2\pi}\frac{d\varphi}{B_0^2}\right)^{-1}
\int_0^{2\pi} \frac{d\varphi}{B_0^4} \left[-\frac{3}{4}(B_{1s}^2+B_{1c}^2)+  B_{20}B_0\right]
+\frac{\mu_0 r^2 p_2}{\pi} \int_0^{2\pi}\frac{d\varphi}{B_0^2}.
\label{eq:W_general}
\end{align}
This result applies to a general toroidal plasma.
In quasisymmetry, (\ref{eq:W_general}) reduces to
\begin{align}
W = 2 r^2\left( -\frac{3}{4}\bar{\eta}^2 +  \frac{\bar{B}_{20}}{B_0} + \frac{\mu_0  p_2}{B_0^2}\right).
\label{eq:W_QS}
\end{align}


\subsection{Numerical verification}

\begin{figure}
  \centering
\includegraphics[width=5.0in]{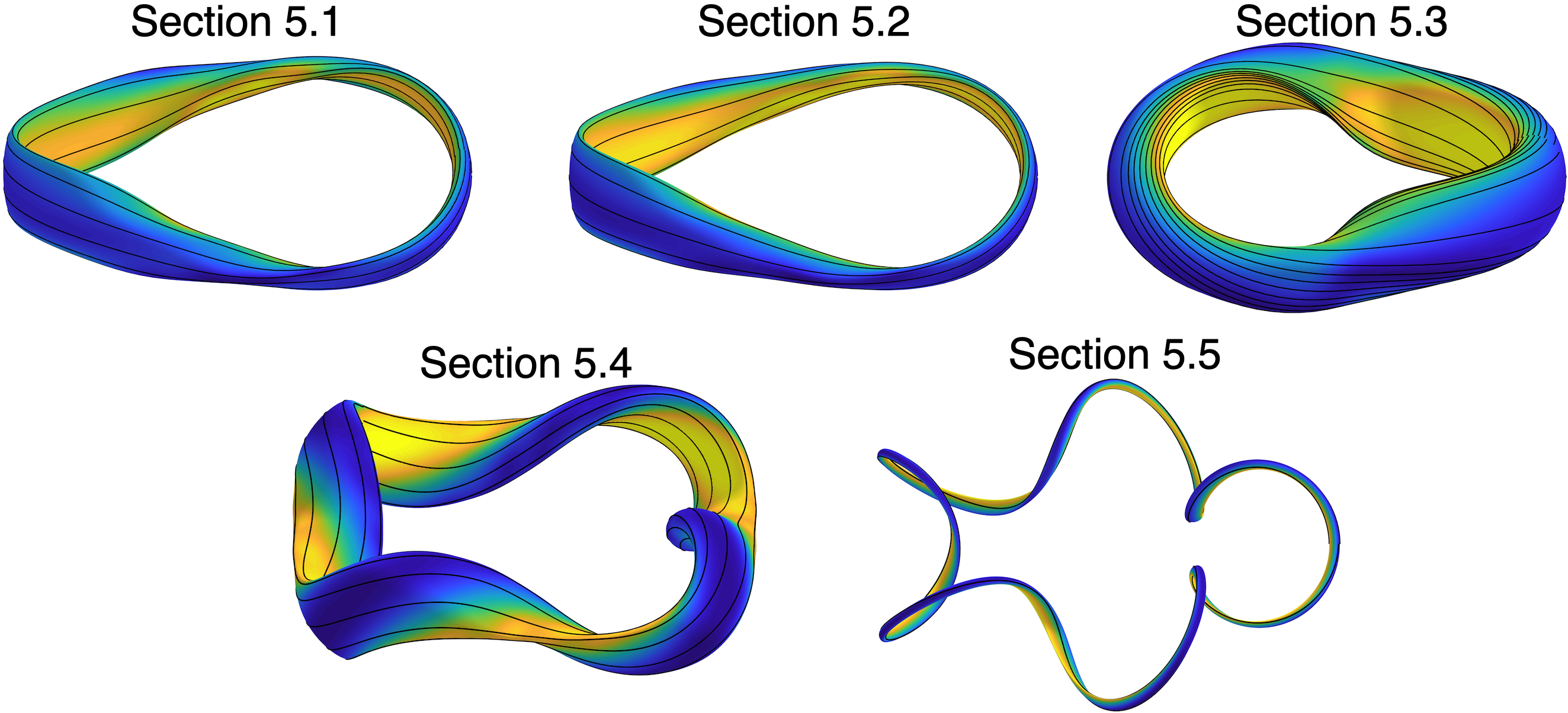}
  \caption{
  \changed{
The five magnetic configurations considered in this paper. Section numbers refer to \cite{r2GarrenBoozer},
where the axis shape and other parameters of each configuration are detailed.
Color indicates the magnetic field strength, and black curves indicate field lines.}
}
\label{fig:5configs}
\end{figure}

\begin{figure}
  \centering
\includegraphics[width=4.5in]{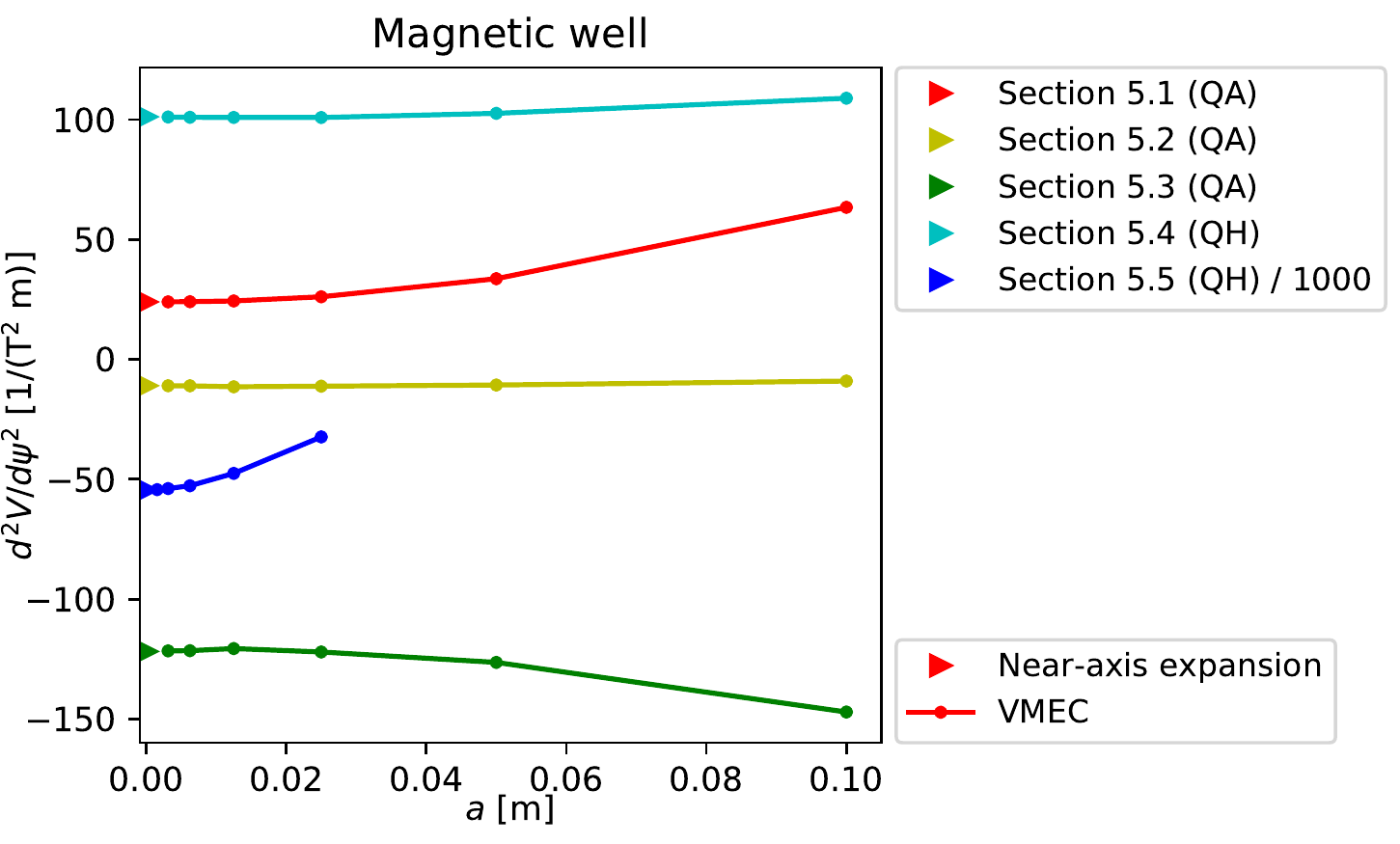}
  \caption{
Verification of the magnetic well calculation.
The five colors indicate the five quasisymmetric configurations detailed in section 5 of \cite{r2GarrenBoozer}\changed{, also shown in figure \ref{fig:5configs}}. Triangles show $d^2 V / d \psi^2$ evaluated from (\ref{eq:wellQS}).
Points connected by lines show $d^2 V / d \psi^2$
computed on the magnetic axis of VMEC configurations
constructed with various values of the boundary effective minor radius $a$, all with mean major axis radius $R = 1$ m. As $a/R \to 0$, the VMEC results converge to (\ref{eq:wellQS}), verifying the calculations.
}
\label{fig:magneticWellVerification}
\end{figure}

The preceding near-axis approach to evaluating magnetic well
can be compared to the magnetic well of finite-aspect-ratio
MHD solutions. Figure \ref{fig:magneticWellVerification} shows
such a comparison for five families of magnetic configuration,
the five cases considered in section 5 of \cite{r2GarrenBoozer}.
\changed{For convenience, these configurations are shown in figure \ref{fig:5configs}.}
This set includes both quasi-axisymmetric and quasi-helically symmetric configurations,
and includes both vacuum fields and configurations
with plasma  pressure and current.
Each family is based upon a single solution of the near-axis equations. 
Substituting several finite values $a$ into the effective minor radius $r$ then yields a set of boundary toroidal surfaces, which
are each provided as input to fixed-boundary MHD solutions using the VMEC code \citep{VMEC1983}.
In figure \ref{fig:magneticWellVerification}, triangles
show the  magnetic well evaluated from eq (\ref{eq:wellQS}) for the underlying near-axis solution,
while dots show $d^2 V / d \psi^2$ evaluated on the magnetic axis from the corresponding VMEC solutions.
For the example of section 5.5, values are divided by 100 in order to fit on the same axes.
For all five examples, as the boundary minor radius $a$ is reduced, the VMEC
results converge to the near-axis results as desired.
The code and data for these calculations
can be obtained online \citep{zenodoCode,zenodoData}.

\section{Mercier Criterion}
\label{sec:Mercier}

The Mercier criterion \citep{Mercier1964,MercierLuc} is a geometrical quantity that, in the context of ideal MHD, allows us to assess the stability of the plasma to radially localized perturbations around a rational surface.
Using the present notation, the criterion can be written as
$\DMerc>0$ where
\begin{align}
    (2\pi)^6 \DMerc=\left[s_G 2\pi^2\frac{d\iota}{d\psi}-\int \frac{\vect{ B} \cdot \vect{\Xi}}{|\nabla \psi|^3}dS\right]^2+\left[s_\psi \mu_0 \frac{dp}{d\psi}\frac{d^2 V}{d\psi^2}-\int \frac{|\vect{\Xi}|^2 dS}{|\nabla \psi|^3}\right]\int \frac{B^2 dS}{|\nabla \psi|^3},
\label{eq:mercric}
\end{align}
and $\vect{\Xi}=\mu_0 \vect{ J} - I'(\psi) \vect{ B}$.
The coefficients $s_G=\pm 1$ and $s_\psi=\pm 1$ correspond to the signs of $G$ and $\psi$, respectively, and ensure invariance under certain changes of coordinates (see \cref{app:mercritsigns}).
We note that $\DMerc$ corresponds to Eq. (37) in \cite{MercierLuc}
multiplied by $\iota^6$, and the integrals are performed along a surface of constant $\psi$ such that $dS=|\nabla\psi||\sqrt{g}|d\vartheta d\varphi$ with $\sqrt{g}$ the Jacobian in $(\psi, \vartheta, \varphi)$ coordinates.

Our goal is to evaluate \cref{eq:mercric} using the near-axis expansion based on the Garren-Boozer formalism.
We focus on the leading order terms of the Mercier criterion, which in our expansion amounts to computing the $O(\epsilon^{-2})$ component for each term in \cref{eq:mercric}.
As shown below, a comparison with numerical results is made using an equivalent form of the Mercier criterion appearing in \cite{Bauer1984,Ichiguchi}.


\subsection{Computation of the criterion at lowest order}

We start by ordering the terms appearing in \cref{eq:mercric}.
In the first brackets, the term proportional to the magnetic shear $d\iota/d\psi$ is $O(\epsilon^0)$.
The term proportional to $\vect{ B} \cdot \vect{\Xi}$ is also $O(\epsilon^0)$. To show this we first note
from (\ref{eq:Boozer_h}) that
\begin{align}
\label{eq:BDotCurlB}
\mu_0\vect{J}\cdot\vect{B} = \left( 
\frac{G}{r\bar{B}}\frac{dI}{dr} - \frac{I}{r\bar{B}}\frac{dG}{dr}
-(G+IN)\frac{\partial\beta}{\partial\vartheta} + I \frac{\partial\beta}{\partial\varphi} \right)
\frac{B^2}{G+\iota I}.
\end{align}
Using this result we obtain
\begin{equation}
    \int \frac{\vect{ B} \cdot \vect{\Xi}}{|\nabla \psi|^3}dS\simeq \int \frac{\sqrt g d\vartheta d\varphi}{|\nabla \psi|^2}[r B_0^2 \left(\beta_{1c}\sin\vartheta - \beta_{1s}\cos\vartheta\right)+O(\epsilon^2)].
\label{eq:Dcurr}
\end{equation}
To evaluate the denominator, 
$\nabla\psi=\sqrt{g}^{-1}\partial\vect{r}/\partial\vartheta \times \partial\vect{r}/\partial\varphi$ with (\ref{eq:positionVector})-(\ref{eq:radial_expansion}) gives
\begin{align}
\label{eq:gradpsi2}
|\nabla\psi|^2 \simeq \frac{r^2 B_0^2 V_1}{2} \left( 1 + a \cos 2\vartheta + b \sin 2\vartheta\right),
\end{align}
where
\begin{align}
\label{eq:V1}
V_1 = &X_{1s}^2 + X_{1c}^2 + Y_{1s}^2 + Y_{1c}^2, \\
a = &( X_{1s}^2 - X_{1c}^2 + Y_{1s}^2 - Y_{1c}^2) / V_1, \\
b = & -2(X_{1s} X_{1c} + Y_{1s} Y_{1c}) / V_1.
\label{eq:b}
\end{align}
Therefore $|\nabla \psi|^2\simeq O(\epsilon^2)$, and the overall $O(\epsilon^{-1})$ contribution to (\ref{eq:Dcurr}) vanishes upon integration over $\vartheta$, leaving the $\vect{B}\cdot\vect{\Xi}$ term in \cref{eq:mercric} as $O(\epsilon^0)$.

Next, we evaluate the term in \cref{eq:mercric} proportional to $|\vect{\Xi}|^2$.
Using $|\vect{ J}|^2=(\vect{ J} \cdot \vect{B})^2/B^2+(dp/d\psi)^2|\nabla \psi|^2/B^2$, we obtain
\begin{equation}
|\vect{\Xi}|^2 = \left(\mu_0 \frac{dp}{d\psi}\right)^2 \frac{|\nabla\psi|^2}{B^2}
+\frac{r^2 B_0^2}{2}\left[ \beta_{1s}^2 + \beta_{1c}^2 + (\beta_{1s}^2-\beta_{1c}^2)\cos 2\vartheta - 2 \beta_{1s} \beta_{1c}\sin 2\vartheta \right] + O(\epsilon^3).
\end{equation}
Carrying out the integrations over $\vartheta$,
the term in (\ref{eq:mercric}) is found to be
\begin{align}
I_{\Xi^2} &= \int \frac{|\vect{\Xi}|^2 dS }{|\nabla\psi|^3}=2\pi |G_0|\left(\frac{4\mu_0^2 p_2^2  }{\bar{B}^2}\int_0^{2\pi}\frac{d\varphi}{B_0^4}+I_\beta\right),
\end{align}
where
\begin{align}
\label{eq:Ibeta_general}
I_\beta = \int_0^{2\pi} \frac{d\varphi}{B_0^2 V_1}\frac{(a^2+b^2)(\beta_{1s}^2+\beta_{1c}^2)+(\sqrt{1-a^2-b^2} - 1)[a(\beta_{1s}^2-\beta_{1c}^2)-2b\beta_{1s} \beta_{1c}]}{\sqrt{1-a^2-b^2}(a^2+b^2)}.
\end{align}
Finally, the integral at the end of (\ref{eq:mercric}) is
\begin{equation}
 \int \frac{B^2\, dS}{|\nabla\psi|^3}
 \approx \frac{2\pi |G_0|}{r^2|\bar B|}\int \frac{d\varphi}{B_0}
 =\frac{2\pi L}{r^2|\bar B|},
\end{equation}
where $V_1 \sqrt{1-a^2-b^2}=2|X_{1s}Y_{1c}-X_{1c}Y_{1s}|=2|\bar{B}|/B_0$ was used, 
and $L=|G_0|\int_0^{2\pi} d\varphi/B_0>0$ is the axis length.
We thereby obtain the following form for the Mercier criterion at lowest order in $\epsilon$:
\begin{equation}
\DMerc =  \frac{\mu_0 p_2 L}{16 \pi^5  r^2 \bar{B}^2}
\left[ \frac{d^2 V}{d\psi^2} - \frac{4 \pi |G_0| \mu_0 p_2}{|\bar{B}|} \int_0^{2\pi} \frac{d \varphi}{B_0^4} - \frac{\pi |G_0 \bar{B}|  I_\beta}{\mu_0 p_2} \right] .
\label{eq:mercricNA}
\end{equation}
In this expression, $V''$ can be evaluated using (\ref{eq:well1_general}).

\cite{Mercier1964} and \cite{MercierLuc} observed that the quantity in the first pair of square brackets in (\ref{eq:mercric}) is smaller than the second near the axis,
consistent with our calculation here. Also, noting that $\beta_1 \propto p_2$ (see (A.52) in \cite{r2GarrenBoozer}), (\ref{eq:mercricNA}) shows that as the pressure gradient becomes small ($p_2 \to 0$), the magnetic well $V''$ becomes the dominant term in $\DMerc$.
(If the limit of a vacuum field is taken before the near-axis limit,
the result is different, $\DMerc \to (d\iota/d\psi)^2/(16 \pi^2)$.)

In the case of quasisymmetry, a number of simplifications are possible. In this case, as shown in Appendix A.3 of \cite{r2GarrenBoozer}, 
$B_0=$ constant, $\bar{B} = s_\psi B_0$,  $|X_{1c}Y_{1s}|=1$, $\beta_{1c}=0$, and
\begin{align}
\beta_{1s} = -\frac{4 s_\psi \mu_0 p_2 G_0 \etabar}{\iota_{N0}B_0^3}.
\end{align}
Therefore (\ref{eq:Ibeta_general}) reduces to
\begin{align}
I_\beta &= \frac{16  \mu_0^2 p_2^2 G_0^2 \etabar^2}{B_0^8 \iota_{N0}^2}
\int_0^{2\pi} d\varphi \frac{X_{1c}^2 +Y_{1c}^2+ 1}{X_{1c}^2 + Y_{1c}^2+Y_{1s}^2 + 2}
\\
&=\frac{16  \mu_0^2 p_2^2 G_0^2 \etabar^2}{B_0^8 \iota_{N0}^2}
\int_0^{2\pi} d\varphi \frac{\etabar^4 + \kappa^4 \sigma^2 + \etabar^2 \kappa^2}{\etabar^4 +\kappa^4(1+\sigma^2) + 2\etabar^2\kappa^2},
\nonumber
\end{align}
where $\sigma = Y_{1c}/Y_{1s}$ is the quantity appearing in eq (A6) of \cite{GB2} and (2.14) of \cite{r2GarrenBoozer}.
Then (\ref{eq:mercricNA}) becomes
\begin{align}
\label{eq:mercricNA_QS}
\DMerc = &\frac{ |G_0| \mu_0 p_2}{8 \pi^4 r^2  B_0^3}
\left[ \frac{d^2 V}{d\psi^2} - \frac{8 \pi^2  |G_0| \mu_0 p_2}{B_0^5} 
\right. \\
& \left. \hspace{0.7in}- \frac{ 16 \pi |G_0|^3 \mu_0 p_2 \etabar^2}{B_0^7 \iota_{N0}^2}
\int_0^{2\pi} d\varphi \frac{\etabar^4 + \kappa^4 \sigma^2 + \etabar^2 \kappa^2}{\etabar^4 +\kappa^4(1+\sigma^2) + 2\etabar^2\kappa^2} \right],
\nonumber
\end{align}
and (\ref{eq:wellQS}) can be applied.


\subsection{Alternative form of the Mercier criterion}

An equivalent form of the Mercier criterion in \cref{eq:mercric} is given in \cite{Bauer1984,Ichiguchi}:
\begin{align}
\DMerc = \DShear + \DCurr + \DWell + \DGeod > 0,
\label{eq:mercritBauer}
\end{align}
where
\begin{align}
\label{eq:DShear}
\DShear = &\frac{1}{16\pi^2} \left( \frac{d\iota}{d\psi}\right)^2, 
\\
\label{eq:DCurr}
\DCurr = &-\frac{s_G}{(2\pi)^4} \frac{d\iota}{d\psi} \int dS  \frac{\vect{\Xi}\cdot \vect{B}}{|\nabla\psi|^3},
\\
\DWell = &\frac{\mu_0}{(2\pi)^6} \frac{dp}{d\psi}\left( s_\psi
\frac{d^2 V}{d\psi^2} 
- \mu_0\frac{dp}{d\psi}
\int \frac{dS}{B^2 |\nabla \psi|}
\right) \int dS \frac{B^2}{|\nabla\psi|^3},
\label{eq:DWell} \\
\DGeod =& \frac{1}{(2\pi)^6}\left(
\int dS \frac{\mu_0 \vect{J}\cdot\vect{B}}{|\nabla\psi|^3} \right)^2
- \frac{1}{(2\pi)^6}\left( \int dS \frac{B^2}{|\nabla\psi|^3}\right)
\int dS \frac{(\mu_0\vect{J}\cdot\vect{B})^2}{B^2 |\nabla\psi|^3}.
\label{eq:DGeod}
\end{align}
These same quantities are reported by VMEC \citep{VMEC1983}.
(Again, we have included factors of $s_G$ and $s_\psi$, as discussed in the appendix.)
The equivalence between \cref{eq:mercric,eq:mercritBauer} can be shown using the identity $(\vect{J} \cdot \vect{B})^2/B^2 =J^2-p'(\psi)^2|\nabla \psi|^2/B^2$.

We now evaluate each term appearing in \cref{eq:mercritBauer} at lowest order in $\epsilon$.
Recall that the scaling of $\DCurr$ with $\epsilon$ was evaluated following (\ref{eq:Dcurr}).
We note that the $\DShear$ and  $\DCurr$ terms are $O(\epsilon^0)$ while the  $\DWell$ and $\DGeod$ terms are $O(\epsilon^{-2})$ so we only need to evaluate the latter two.
These are given by
\begin{equation}
\DWell = \frac{ \mu_0 p_2 L}{16 \pi^5 r^2 \bar{B}^2}
\left[  \frac{d^2 V}{d\psi^2} - \frac{4 \pi\mu_0 p_2 |G_0|}{ |\bar{B}|} \int_0^{2\pi} \frac{d\varphi}{B_0^4}\right] 
\label{eq:DWell_result}
\end{equation}
and
\begin{equation}
\DGeod =  -\frac{|G_0| L  I_\beta}{16 \pi^4  r^2 |\bar{B}|},
 \label{eq:DGeod_result}
\end{equation}
with $I_\beta$ given by (\ref{eq:Ibeta_general}).
To obtain $\DGeod$, it is convenient to subtract (\ref{eq:DWell_result}) from (\ref{eq:mercricNA}).

In the case of quasisymmetry, these last expressions reduce to
\begin{align}
\DWell &= \frac{\mu_0 p_2 |G_0|}{8 \pi^4 r^2 B_0^3} \left[  \frac{d^2 V}{d\psi^2} - \frac{8 \pi^2 \mu_0 p_2 |G_0|}{B_0^5}\right],
\label{eq:DWell_QS}
\\
\DGeod &= -
  \frac{2   \mu_0^2 p_2^2  G_0^4  \etabar^2}{ \pi^3 r^2  B_0^{10}  \iota_{N0}^2}
\int_0^{2\pi} d\varphi \frac{\etabar^4 + \kappa^4 \sigma^2 + \etabar^2 \kappa^2}{\etabar^4 +\kappa^4(1+\sigma^2) + 2\etabar^2\kappa^2} .
\label{eq:DGeod_QS}
\end{align}


\subsection{Numerical Verification}

We now verify the analytic results of the previous section by comparing them to computations using the VMEC equilibrium code. As with figure \ref{fig:magneticWellVerification}, we first compute a numerical solution of the near-axis equations to $O(\epsilon^2)$, then use the procedure of \cite{r2GarrenBoozer} to construct a magnetic surface surrounding this axis for $r$ equal to a small finite value $a$. This surface is then used as the prescribed boundary for a fixed boundary VMEC calculation, with 
\changed{uniform toroidal current density and with pressure profile $p=p_0 + r^2 p_2$ for $p_0 = -a^2 p_2$.
(The configuration remains a solution of the MHD equilibrium equations if any constant is added to $p$, 
as one would do if this configuration were considered to be the core of a lower-aspect-ratio one.)
}
VMEC reports all the individual quantities in (\ref{eq:mercritBauer}) (with the same normalization used here) as a standard diagnostic. VMEC's calculation of $\DGeod$ converges extremely slowly with the number of radial surfaces {\ttfamily NS}, so all calculations shown here use {\ttfamily NS} $\ge 801$, and results from the innermost VMEC grid points are dropped.
The code and data for these calculations
can be obtained online \citep{zenodoCode,zenodoData}.

\changed{
For a first verification scenario we use the
quasi-helically symmetric configuration of section 5.4 of \cite{r2GarrenBoozer}, with two minor modifications. The field strength $B_0$ has been doubled to 2 T to provide test coverage for the $B_0$ factors in the analytic expressions. Also 
a nonzero $p_2 = -2$ MPa/m$^2$ is introduced since otherwise $\DMerc=0$. The complete set of near-axis parameters is therefore given by the axis shape
\begin{align}
R_0(\phi)  \;\mathrm{[m]}=& 1 + 0.1700 \cos(4\phi) + 0.01804 \cos(8\phi) + 0.001409 \cos(12\phi) + 0.00005877 \cos(16\phi), \nonumber\\
z_0(\phi)  \;\mathrm{[m]}= & \hspace{0.25in}0.1583 \sin(4\phi) + 0.01820 \sin(8\phi) + 0.001548 \sin(12\phi) + 0.00007772 \sin(16\phi),
\label{eq:QH_axis_shape}
\end{align}
with 
$\etabar=1.569$ m$^{-1}$, $\sigma(0)=0$, $I_2=0$, $B_{2c}=0.2696$ T/m$^2$, and $B_{2s}=0$.
}

\begin{figure}
  \centering
    \includegraphics[width=5.0in]{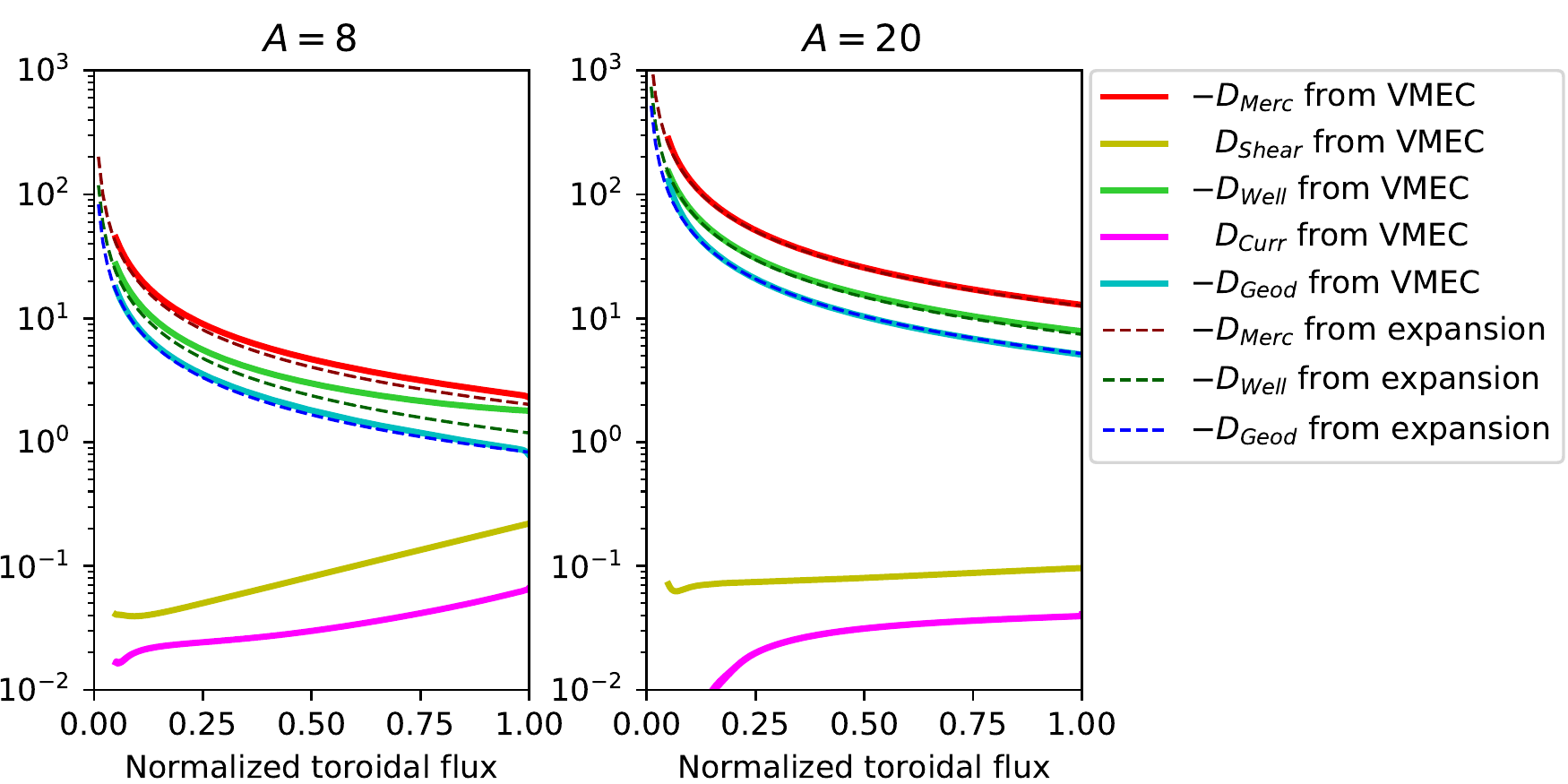}
  \caption{
Numerical verification of (\ref{eq:mercricNA_QS}), (\ref{eq:DWell_QS}), and (\ref{eq:DGeod_QS}) for the terms in Mercier's criterion, by comparison to the VMEC code. The magnetic configuration is the
\changed{quasi-helically symmetric example in section 5.4 of \cite{r2GarrenBoozer},
also described near (\ref{eq:QH_axis_shape}),
with boundary minor radius (a) $a=0.125$ m or (b) $a=0.05$ m.}
}
\label{fig:compareMercierTerms}
\end{figure}

\changed{
Figure \ref{fig:compareMercierTerms} shows 
a comparison of VMEC's evaluation of (\ref{eq:mercritBauer}) to the analytic expressions for this quasi-helically symmetric configuration.
For this figure, results are shown for $A=8$ and $A=20$, where $A=R/a$ is the boundary aspect ratio, and} $R =$ 1 m is the mean of the axis major radius over the standard toroidal angle. As predicted by theory, $\DShear$ and $\DCurr$ are far smaller than the other terms in Mercier's criterion.
Equations (\ref{eq:mercricNA_QS}) and (\ref{eq:DWell_QS}) are evaluated using (\ref{eq:wellQS}).
\changed{At $A=20$,}
excellent agreement is seen between the asymptotic expressions (\ref{eq:mercricNA_QS}), (\ref{eq:DWell_QS}), and (\ref{eq:DGeod_QS}) and VMEC's finite-aspect-ratio evaluations of $\DMerc$, $\DWell$, and $\DGeod$. 
\changed{At $A=8$, the agreement is naturally not as good, but the asymptotic formulae remain qualitatively correct.}

\begin{figure}
  \centering
  \includegraphics[width=4.5in]{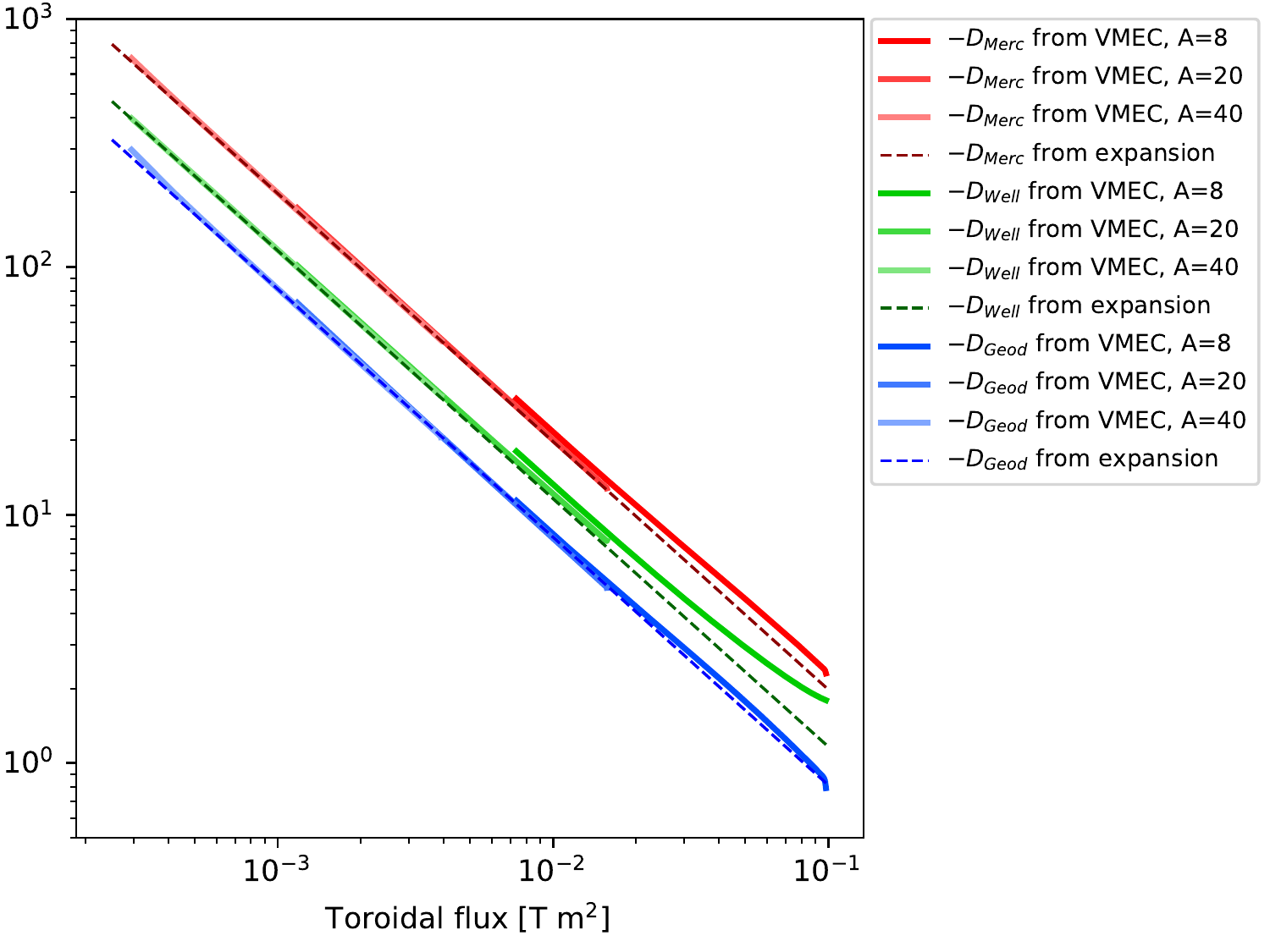}
  \caption{
Numerical verification of equations (\ref{eq:mercricNA_QS}), (\ref{eq:DWell_QS}), and (\ref{eq:DGeod_QS}) for the terms in Mercier's criterion, by comparison to the VMEC code. The magnetic configuration is the
\changed{quasi-helically symmetric example in section 5.4 of \cite{r2GarrenBoozer} and (\ref{eq:QH_axis_shape}).}
}
\label{fig:compareMercierTerms_multi}
\end{figure}

Figure \ref{fig:compareMercierTerms_multi} shows the same comparison repeated for several values of boundary aspect ratio.
Again the agreement between the asymptotic expressions and VMEC calculations is excellent, particularly as $A$ increases. The quantities in the core of the \changed{$A=8$} configuration can be seen to not exactly overlap those from the $A=20$ configuration. This is because the boundary surface is constructed by extrapolating out from the axis \emph{approximately}, but then VMEC computes the equilibrium inside that boundary without a near-axis approximation, leading to a slightly different axis shape than the original one.

\begin{figure}
  \centering
    \includegraphics[width=4.5in]{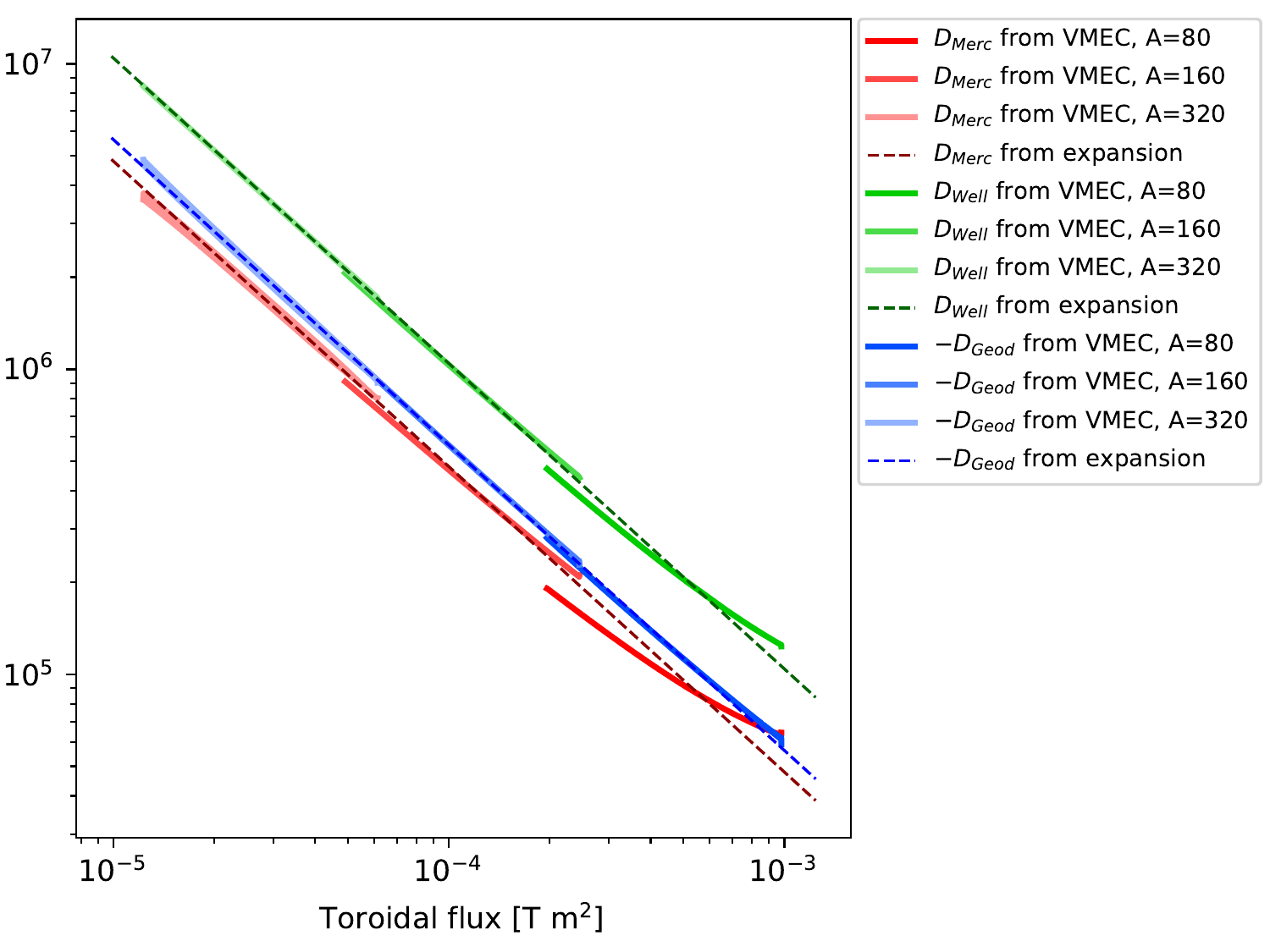}
  \caption{
Numerical verification of equations (\ref{eq:mercricNA_QS}), (\ref{eq:DWell_QS}), and (\ref{eq:DGeod_QS}) for the terms in Mercier's criterion, by comparison to the VMEC code. The magnetic configuration is the quasi-helically symmetric example in section 5.5 of \cite{r2GarrenBoozer} \changed{and (\ref{eq:allTerms_axisShape})}.
}
\label{fig:compareMercierTerms_allTerms}
\end{figure}

Figure \ref{fig:compareMercierTerms_allTerms} presents a similar comparison to
figure \ref{fig:compareMercierTerms_multi}, but now using the example quasi-helically symmetric configuration of section 5.5 of \cite{r2GarrenBoozer}.
This configuration provides a comprehensive test covering all effects,
including departure from stellarator symmetry,
and nonzero quasisymmetry number $N$, pressure $p_2$, and plasma current $I_2$. The field strength $B_0$ has once more been doubled to 2 T to provide test coverage for the $B_0$ factors in the analytic expressions.
\changed{Thus, the complete set of parameters is given by the axis shape
\begin{align}
\label{eq:allTerms_axisShape}
R_0(\phi)  \;\mathrm{[m]}=& 1 + 0.3 \cos(5\phi), \\
z_0(\phi)  \;\mathrm{[m]}= & \hspace{0.25in}0.3 \sin(5\phi),\nonumber
\end{align}
with $\etabar=2.5$ m$^{-1}$, $\sigma(0)=0.3$, $I_2=3.2$ T$/$m, $p_2=-2\times 10^7$ Pa$/$m$^2$, $B_{2c}=2$ T$/$m$^2$, and $B_{2s}=6$ T$/$m$^2$.
}
This configuration is limited to quite high aspect ratio due to large $O(\epsilon^2)$ coefficients, making the expansion only accurate at quite small $r$. Again, the VMEC results are seen to converge to the expressions (\ref{eq:mercricNA_QS}), (\ref{eq:DWell_QS}), and (\ref{eq:DGeod_QS}) as the boundary aspect ratio increases.

\begin{figure}
  \centering
  \includegraphics[width=4.5in]{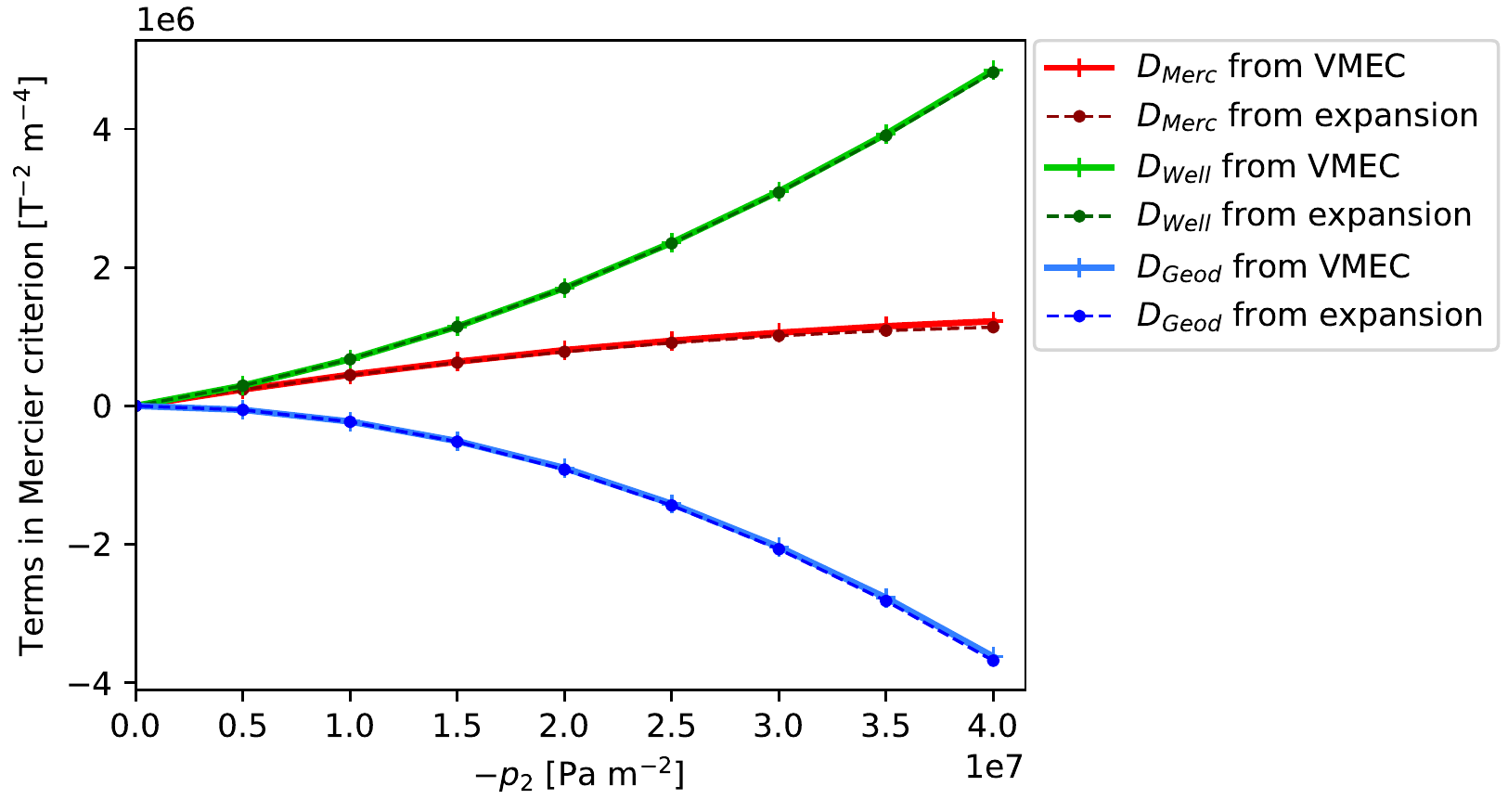}
  \caption{
Numerical verification of equations (\ref{eq:mercricNA_QS}), (\ref{eq:DWell_QS}), and (\ref{eq:DGeod_QS}) for the terms in Mercier's criterion, by comparison to the VMEC code, showing a scan in pressure. The magnetic configuration is the quasi-helically symmetric example in section 5.5 of \cite{r2GarrenBoozer} \changed{and (\ref{eq:allTerms_axisShape})}, for $A=320$.
On the horizontal axis, $p_2$ is the quadratic-in-$r$ term in the pressure profile, and the on-axis pressure is $-p_2 a^2$.
}
\label{fig:pressureScan}
\end{figure}

\begin{figure}
  \centering
  \includegraphics[width=4.5in]{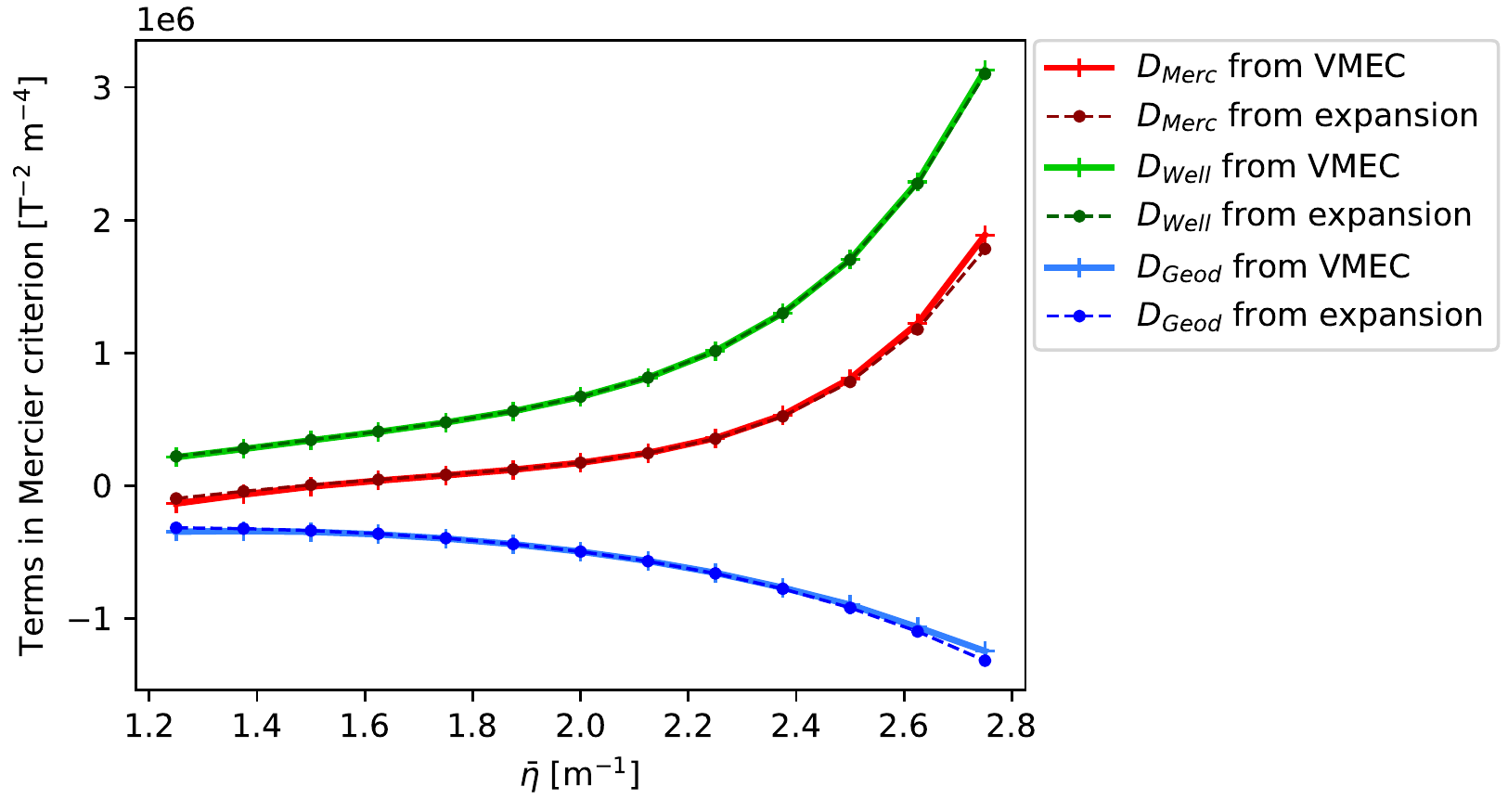}
  \caption{
Numerical verification of equations (\ref{eq:mercricNA_QS}), (\ref{eq:DWell_QS}), and (\ref{eq:DGeod_QS}) for the terms in Mercier's criterion, by comparison to the VMEC code, showing a scan in the input parameter $\etabar$. The magnetic configuration is the quasi-helically symmetric example in section 5.5 of \cite{r2GarrenBoozer} \changed{and (\ref{eq:allTerms_axisShape})}, for $A=320$.
}
\label{fig:etaBarScan}
\end{figure}

Finally, figures \ref{fig:pressureScan}-\ref{fig:etaBarScan}
present comparisons between our analytic expressions and VMEC's results as we scan the input parameters $p_2$ and $\etabar$ respectively. For these figures, the VMEC results are taken from the outermost VMEC grid point.
Good agreement is seen across both parameter scans, providing comprehensive verification.

\section{Necessary condition for resistive MHD stability}
\label{sec:resistive}


We next consider whether the stability of radially localized interchange modes near the axis is substantially different for resistive MHD compared to ideal MHD.
In \cite{Glasser1975}, assuming $\DMerc>0$, a necessary condition for the plasma stability under the conditions with which the resistive MHD equations are valid was derived.
The criterion is given by
\begin{equation}
    D_R = E+F+4 \pi^2 (\iota')^{-2} H^2 \le 0
\label{eq:mercresis}
\end{equation}
where
\begin{align}
\label{eq:GlasserE}
    E=& \frac{1}{(2\pi)^6} \left[\Lambda \frac{\int \mu_0 \vect{ J} \cdot \vect{ B} dS/|\nabla \psi|}{\int B^2 dS/|\nabla \psi|}\int \frac{B^2 dS}{|\nabla \psi|^3}-\left(s_\psi V'' \mu_0 p'+\Lambda I'\right)\int \frac{B^2 dS}{|\nabla \psi|^3}\right],
    \\
    F=&\frac{1}{(2\pi)^6} \left[\int \frac{B^2 dS}{|\nabla \psi|^3} \int \frac{(\mu_0\vect{ J} \cdot \vect{ B})^2}{B^2|\nabla \psi|^3} dS \right. \\
    & \hspace{0.6in} \left.+ \mu_0^2 p'^2 \int \frac{B^2 dS}{|\nabla \psi|^3}\int \frac{dS}{B^2|\nabla \psi|} - \left(\int \frac{\mu_0 \vect{ J} \cdot \vect{ B}}{|\nabla \psi|^3} dS \right)^2 \right], \nonumber
    \\
    H=&\frac{\Lambda}{(2\pi)^6}
    \left(\int \frac{\mu_0 \vect{ J} \cdot \vect{ B}}{|\nabla \psi|^3} dS- \int \frac{B^2 dS}{|\nabla \psi|^3}\frac{\int \mu_0 \vect{ J} \cdot \vect{ B} dS/|\nabla \psi|}{\int B^2 dS/|\nabla \psi|}\right),
    \label{eq:GlasserH}
\end{align}
and $\Lambda = s_G 4 \pi^2 \iota'$.
Primes denote $d/d\psi$.
We note that \cref{eq:mercresis} corresponds to Eq. (16) of \cite{Glasser1975} multiplied by $(\iota')^2 /(4 \pi^2)$.

We now show that, to lowest order in $\epsilon$, i.e., at $O(\epsilon^{-2})$, $D_R \simeq -\DMerc$.
We first note that the criterion in \cref{eq:mercresis} can be related to the Mercier criterion $\DMerc$ by rewriting $\DMerc$ as
\begin{equation}
    \DMerc = \frac{(\iota')^2}{16 \pi^2}-E-F-H>0.
\label{eq:mercgla}
\end{equation}
The functions $D_R$ and $\DMerc$ are then related via
\begin{equation}
    D_R=-\DMerc+\frac{4 \pi^2}{(\iota')^2} \left[ H -\frac{(\iota')^2}{8\pi^2}\right]^2.
\end{equation}
We have already seen that $\iota'(\psi)$ is $O(\epsilon^0)$.
Using (\ref{eq:BDotCurlB}), the function $H$ turns out to have a rather compact form when written in terms of Boozer coordinates:
\begin{equation}
    H=\frac{\iota'}{(2\pi)^4}  \int d\vartheta d\varphi \frac{I \partial\beta/\partial \varphi - (G+IN) \partial \beta/\partial \vartheta}{|\nabla \psi|^2}.
\label{eq:Hbooz}
\end{equation}
The $\epsilon \ll 1$ expansion has not yet been employed.
As the integral over $\vartheta$ of the $O(\epsilon^{-1})$ component of \cref{eq:Hbooz} vanishes and $I=O(\epsilon^2)$, we conclude that $H=O(\epsilon^0)$ and thus $\DMerc\simeq -D_R+O(\epsilon^0)$.
Therefore, the distinction between ideal and resistive interchange stability (for the radially localized modes of Mercier and Glasser) becomes insignificant near the axis.
A special case of this result, for circular-cross-section axisymmetric equilibria, was noted in \cite{Glasser1976}.


\section{Discussion and conclusions}

In the analysis above we have shown that the magnetic well
and Mercier stability can be computed directly from a solution of the near-axis equilibrium equations in Garren \& Boozer's expansion. 
As demonstrated by the agreement between our analytic formulae and finite-aspect-ratio
VMEC calculations in the figures, it is therefore possible to assess the Mercier
stability of the constructed configurations without running a finite-aspect-ratio equilibrium code.
These results contribute to the goal of
carrying out stellarator design in part within the near-axis approximation, which may help resolve the problem of numerical optimizations getting stuck in local minima, as follows. In the near-axis approximation, the equilibrium equations can be solved many orders of magnitude faster than the full 3D equilibrium equations, enabling global surveys of the landscape of possible configurations that are not limited to the vicinity of a single optimum. In such a survey, the results in the present paper let us immediately exclude Mercier-unstable configurations. 
The most promising configurations from a high-aspect-ratio survey 
can then be provided as new initial conditions for traditional local optimization with a finite-aspect-ratio 3D equilibrium code.

The key results from our analysis are as follows.
Near the magnetic axis, the magnetic well is given by (\ref{eq:well1_general}), (\ref{eq:What_nearAxis}), and (\ref{eq:W_general}), depending on the definition used.
In the special case of quasisymmetry, these expressions simplify to (\ref{eq:wellQS}), (\ref{eq:What_QS}), and (\ref{eq:W_QS}).
In Mercier's criterion,
the terms
$\DShear$ and $\DCurr$ scale as $\propto r^0$, while the terms $\DWell$ and $\DGeod$ scale as $r^{-2}$.
Therefore $\DShear$ and $\DCurr$ are negligible near the axis, and overall $\DMerc\propto r^{-2}$. The dominant terms near the axis can be computed from (\ref{eq:mercricNA}), (\ref{eq:DWell_result}), and (\ref{eq:DGeod_result}), with (\ref{eq:Ibeta_general}).
In the case of quasisymmetry,
these expressions simplify to
(\ref{eq:mercricNA_QS}), (\ref{eq:DWell_QS}), and (\ref{eq:DGeod_QS}).
Finally, Glasser's criterion for resistive interchange stability coincides with Mercier's ideal criterion to leading order near the axis.

\changed{
The dependence of these formulae on pressure must be interpreted carefully.
If specific coils are considered fixed, then a change to the plasma pressure profile would affect the magnetic well and Mercier stability not only through the explicit $p_2$ terms in the expressions here. In such a free-boundary scenario the axis shape would also generally change, causing changes to $B_{1c}$, $B_{1s}$, $B_{20}$, and other quantities in these formulae.
}

\changed{
One question arising from this work is whether it is possible to calculate the maximum minor radius at which the expressions in this paper are accurate.
An estimate of this bound could perhaps be derived by computing the next terms in the $\epsilon \ll 1$ expansion of Mercier's criterion, and identifying the value of $r$ at which these corrections are comparable to the leading-order terms. 
Calculation of these correction terms would require substantial work, and we will not attempt to do so here.
A cruder estimate of the maximum accurate $r$ can be obtained from the ratio of the surface shape functions $(X_1,Y_1)$ compared to $(X_2,Y_2)$. One meaningful combination of these quantities is the value of $r$ at which the surface shapes would acquire a cusp or other singularity if the expansions of $(X,Y,Z)$ are truncated after $(X_2,Y_2,Z_2)$.
This calculation will be presented in a separate publication.
}


We acknowledge valuable conversations about this research with Sophia Henneberg, Chris Hegna, John Schmitt, and Alan Glasser.
This work was supported by the
U.S. Department of Energy, Office of Science, Office of Fusion Energy Science,
under award number DE-FG02-93ER54197.
This work was also supported by a grant from the Simons Foundation (560651, ML).

\appendix


\section{Signs in Mercier's criterion}
\label{app:mercritsigns}

Typically two possible transformations exist in which the signs of certain flux coordinates are flipped while physical quantities such as $\vect{B}$ are unchanged. Some expressions for Mercier stability in the literature are not invariant under these transformations, due to assumptions about signs made during their derivation. Here we show how to generalize these forms of Mercier's criterion so they are invariant.

\subsection{Parity transformations}

Let us first precisely state the two coordinate transformations under which physical phenomena should be invariant. The magnetic field is represented in Boozer coordinates as
\begin{align}
\label{eq:B_straightFieldLines}
\vect{B} &= \frac{1}{2\pi}\left(\nabla\Psi\times\nabla\theta + \nabla\varphi\times\nabla\Phi\right)
\\
&= \frac{\beta}{2\pi} \nabla\Psi + I(\Psi) \nabla\theta + G(\Psi) \nabla\varphi,
\label{eq:B_Boozer}
\end{align}
where $\Psi$ and $\Phi$ are the toroidal and poloidal fluxes (not divided by $2\pi$), and the angles $\theta$ and $\varphi$ are periodic with period $2\pi$. The rotational transform is $\iota = d\Phi / d \Psi$.

In ``parity transformation 1,'' the signs of $\Psi$, $\theta$, $\beta$, $I$, and $\iota$ are flipped, while $\varphi$, $G$, and $\Phi$ are unchanged. This transformation leaves $\vect{B}$ unchanged in both (\ref{eq:B_straightFieldLines}) and (\ref{eq:B_Boozer}). In ``parity transformation 2,'' the signs of $\varphi$, $G$, $\Phi$, and $\iota$ are flipped, while 
$\Psi$, $\theta$, $\beta$, and $I$ are unchanged.
This transformation again leaves $\vect{B}$ unchanged in both (\ref{eq:B_straightFieldLines}) and (\ref{eq:B_Boozer}).

Since $\vect{B}$ is unchanged by these transformations, physical consequences such as stability should be unaltered. Expressions for Mercier stability in references such as \cite{Mercier1964,MercierLuc,Bauer1984} however appear to have sign changes in at least some terms.
In the remainder of this section we show how these expressions for Mercier stability should be modified to handle these coordinate transformations.

\subsection{Greene \& Johnson form}

It is convenient to begin with a form of the criterion developed by \cite{GreeneJohnson1961,GreeneJohnson1962,GreeneJohnson1962Err}, since they explicitly state their assumption about the coordinate signs. They use Hamada coordinates $(V,\theta_H,\varphi_H)$, with angles that are periodic with period 1, which are required to form a right-handed system:
\begin{align}
\nabla V \cdot \nabla \theta_H \times \nabla \varphi_H = +1.
\label{eq:Hamada}
\end{align}
We require the flux surface volume $V$ to be $\ge 0$. The magnetic field satisfies
\begin{align}
\label{eq:Hamada_straightFieldLines}
\vect{B} = \nabla\Psi_H \times \nabla \theta_H + \nabla\varphi_H \times\nabla\Phi_H,
\end{align}
where we have included the $H$ subscript to emphasize that the toroidal and poloidal fluxes $\Psi_H$ and $\Phi_H$ must have signs compatible with those of $(\theta_H,\varphi_H)$ in (\ref{eq:Hamada_straightFieldLines}), which must in turn be consistent with (\ref{eq:Hamada}). Note $\Psi = \pm \Psi_H$ and $\Phi = \pm \Phi_H$ are possible. Neither of the two parity transformations is allowed by itself in Greene \& Johnson's coordinates (e.g. we cannot replace $(\Psi_H, \theta_H) \to (-\Psi_H, -\theta_H)$) since (\ref{eq:Hamada}) would be violated. However applying both transformations simultaneously is allowed.

The stability criterion in \cite{GreeneJohnson1961}, also  eq (40) in \cite{GreeneJohnson1962}, is $D_{GJ}>0$ where
\begin{align}
\label{eq:GreeneJohnson}
D_{GJ} = &\left[\oint \frac{B\, d\ell}{|\nabla V|^2}\right]^{-1}
\left[ \left( \frac{d \Psi_H}{dV} \frac{d^2 \Phi_H}{dV^2}
-\frac{d^2 \Psi_H}{dV^2} \frac{d\Phi_H}{dV}\right) \oint \frac{d\ell}{2B} 
-\oint \frac{\mu_0\vect{J}\cdot\vect{B}\, d\ell}{B |\nabla V|^2}\right]^2 
\\
&-2 \oint \frac{d\ell \, \mu_0\vect{J}\times\nabla V \cdot (\vect{B}\cdot\nabla)\nabla V}{B |\nabla V|^4},
\nonumber
\end{align}
and here $\ell$ denotes arclength along a closed field line on the rational surface.
(\cite{GreeneJohnson1961}, \cite{Mercier1964}, \cite{Bauer1984}, \cite{Ichiguchi}, and \cite{Glasser1975} all set $\mu_0 \to 1$; we restore $\mu_0$ by replacing $\vect{J} \to \mu_0 \vect{J}$ and $p \to \mu_0 p$.)
Note the factor of $-2$ in the last term of (\ref{eq:GreeneJohnson}) is missing in \cite{GreeneJohnson1962}, as noted in \cite{GreeneJohnson1962Err}, but correct in \cite{GreeneJohnson1961}.

In our Boozer coordinates (\ref{eq:B_straightFieldLines})-(\ref{eq:B_Boozer}), the quantity
\begin{align}
\nabla V\cdot\nabla\theta\times\nabla\varphi = \frac{dV}{dr} \frac{B^2}{(G+\iota I)r\bar{B}}
=s_G s_\psi \left| \frac{dV}{dr} \frac{B^2}{(G+\iota I) r\bar{B}} \right|
\end{align}
could have either sign. Here, $s_\psi = \sgn(\Psi)=\sgn(\bar{B})=\pm 1$ and $s_G=\sgn(G)=\pm 1$,
and we have used $dV/dr>0$.
Therefore we cannot necessarily take the poloidal and toroidal fluxes from the Boozer coordinate system and insert them in (\ref{eq:GreeneJohnson}), setting $\Psi_H=\Psi$  and $\Phi_H=\Phi$. This is only allowed if  $s_G s_\psi = +1$. If  $s_G s_\psi = -1$, we must perform transformation 1 or 2 (not both) before substituting the fluxes into (\ref{eq:GreeneJohnson}). In this case, we would have either $(\Psi_H,\Phi_H)=(-\Psi,\Phi)$
or $(\Psi,-\Phi)$
This effect can be achieved by inserting a factor $s_G s_\psi$ in Greene \& Johnson's expression:
\begin{align}
\label{eq:GJ_generalized}
D_{GJ} = &\left[\oint \frac{B\, d\ell}{|\nabla V|^2}\right]^{-1}
\left[ 
\frac{s_G s_\psi}{2}
\left( \frac{d\Psi}{dV}\right)^2 \frac{d\iota}{dV}
\oint \frac{d\ell}{B} 
-\oint \frac{\mu_0\vect{J}\cdot\vect{B}\, d\ell}{B |\nabla V|^2}\right]^2 
\\
&-2 \oint \frac{d\ell \, \mu_0\vect{J}\times\nabla V \cdot (\vect{B}\cdot\nabla)\nabla V}{B |\nabla V|^4}.
\nonumber
\end{align}
This statement of the stability condition is now in a form invariant under either transformation.

The same argument can be applied to expressions in \cite{Glasser1975}, who use
the same Hamada coordinates with positive Jacobian. By this reasoning, the expressions for $E$ and $H$ in eq (13) of \cite{Glasser1975} are made parity invariant by including a factor $s_G s_\psi$; $F$ does not acquire this factor. The results, multiplied by $(d\iota/d\psi)^2/(4\pi^2)$, give (\ref{eq:GlasserE})-(\ref{eq:GlasserH}).

Following \cite{MercierLuc}, the line integrals in (\ref{eq:GJ_generalized}) can
be approximately converted to area integrals 
by
\begin{align}
\left. \oint \frac{Q\, d\ell}{B} \middle/ \oint \frac{ d\ell}{B} \right.
\approx \left. \int \frac{Q\, dS}{|\nabla \Phi|} \middle/ 
\int \frac{ dS}{|\nabla \Phi|} \right.
\end{align}
where $Q$ is any quantity. Noting $\int dS/|\nabla\Phi| = s_\psi s_\iota dV/d\Phi$ (which follows from (\ref{eq:volume})), we find the stability condition can be written as $D_{M} \ge 0$ where
\begin{align}
\label{eq:Mercier_intermediate}
D_{M} =  \left[
\frac{s_G }{2} \frac{d(1/|\iota|)}{d\Phi} + \int \frac{\mu_0\vect{J}\cdot\vect{B}\, dS}{|\nabla\Phi|^3}
\right]^2
-2
\left[ \int \frac{B^2\, dS}{|\nabla\Phi|^3}\right]
\int \frac{dS\, \mu_0\vect{J}\times\vect{n}\cdot(\vect{B}\cdot\nabla\vect{n})}{|\nabla\Phi|^3}.
\end{align}
Here, Mercier's unit vector $\vect{n}=|\nabla\Phi|^{-1}\nabla\Phi$ has been introduced.
The expression (\ref{eq:Mercier_intermediate}) generalizes eq (35) on page 60 of \cite{MercierLuc} to be properly invariant under the two parity transformations.

\subsection{Mercier's form}

To obtain the form of the stability criterion favored by Mercier, we can follow page 61 of \cite{MercierLuc} (also the appendix of \cite{Mercier1964}), beginning with
\begin{align}
\label{eq:Appendix6}
\frac{2 \vect{J}\times\vect{n}\cdot(\vect{B}\cdot\nabla\vect{n})}{|\nabla\Phi|^2} = \frac{\mu_0 J^2}{|\nabla\Phi|^2} - \vect{J}\cdot\nabla\left( \frac{\vect{n}\cdot\nabla\chi}{|\nabla\Phi|}\right) - \frac{dp}{d\Phi}\nabla\cdot\left( \frac{\vect{n}}{|\nabla\Phi|}\right),
\end{align}
where $\chi = (\varphi - \theta/\iota)/(2\pi)$.
The derivation of this identity by \cite{MercierLuc}
in their appendix 6
makes no assumptions about signs so (\ref{eq:Appendix6}) is already
invariant. Other expressions on
 page 61 of \cite{MercierLuc} however require modification to account for signs, including
\begin{align}
\int \frac{dS}{|\nabla\Phi|}\mu_0\vect{J}\cdot\nabla\left( \frac{\vect{n}\cdot\nabla\chi}{|\nabla\Phi|}\right) = \left[ \frac{d(1/\iota)}{d \Phi}\right] \frac{s_G}{|\iota|} \frac{dG}{d\psi}
\end{align}
and
\begin{align}
\int \frac{dS}{|\nabla\Phi|} \nabla\cdot \left( \frac{\vect{n}}{|\nabla\Phi|}\right) = \frac{s_\iota s_\psi}{\iota^2} \frac{d^2 V}{d \Psi^2} + s_\iota s_\psi \left[ \frac{d(1/\iota)}{d\Phi}\right] \frac{dV}{d\Psi},
\end{align}
where $s_\iota = \sgn(\iota) = \pm 1$.
From eq (2.7) in \cite{r2GarrenBoozer}) we find
\begin{align}
\label{eq:MercierMHD}
\frac{dG}{d\psi} + \iota \frac{dI}{d\psi} = -\frac{s_G s_\psi \mu_0}{4\pi^2} \frac{dp}{d\psi} \frac{dV}{d\psi}
\end{align}
instead of the last equation on page 61 of \cite{MercierLuc}. Combining (\ref{eq:Appendix6})-(\ref{eq:MercierMHD}) gives
\begin{align}
2 \int\frac{dS \mu_0\vect{J}\times\vect{n}\cdot(\vect{B}\cdot\nabla\vect{n})}{|\nabla\Phi|^3} = 
\int \frac{dS\, \mu_0^2 J^2}{|\nabla\Phi|^3}
+s_G \left[ \frac{d(1/|\iota|)}{d\Phi}\right] \frac{dI}{d\psi} - \frac{s_\iota s_\psi \mu_0}{\iota^2}\frac{dp}{d\Phi} \frac{d^2V}{d\Psi^2},
\end{align}
which corrects the first equation on page 62 of \cite{MercierLuc}. Using this result in (\ref{eq:Mercier_intermediate}), we obtain Mercier's form of the stability criterion, $D_{M} \ge 0$ with
\begin{align}
D_{M} =  \left[
\frac{s_G }{2} \frac{d(1/|\iota|)}{d\Phi} + \int \frac{\vect{B}\cdot\vect{\Xi}\, dS}{|\nabla\Phi|^3}
\right]^2
+
\left[ \frac{s_\iota s_\psi \mu_0}{\iota^2} \frac{dp}{d\Phi} \frac{d^2V}{d\Psi^2}
-\int \frac{|\vect{\Xi}|^2\, dS}{|\nabla\Phi|^3}\right]
\int \frac{B^2\, dS}{|\nabla\Phi|^3},
\label{eq:MercierCorrected}
\end{align}
where $\vect{\Xi}=\mu_0\vect{J} - \vect{B} dI/d\psi = \mu_0 (\vect{J} - \vect{B}\, dI_t/d\Psi)$,
and $I_t = 2\pi I / \mu_0$ is the toroidal current.
This result, analogous to (37) of \cite{MercierLuc}, is properly invariant under either parity transformation.


\subsection{Geodesic curvature form}

In more recent publications \citep{Bauer1984} and in the VMEC code,
the stability criterion is expressed in a  different form,
which we now derive. Using $J^2 = (\vect{J}\cdot\vect{B})^2/B^2 + (dp/d\Psi)^2|\nabla\Psi|^2/B^2$
(which follows from the square of $\vect{J}\times\vect{B}=\nabla p$) and defining 
$\DMerc = \iota^6 D_{M}$,
(\ref{eq:MercierCorrected}) may be expressed
as $\DMerc \ge 0$ with
\begin{align}
\label{eq:MercierBauerCorrected}
\DMerc = & \frac{1}{4}\left( \frac{d\iota}{d\Psi}\right)^2
-s_G \frac{d\iota}{d\Psi} \iint \frac{d\theta \, d\varphi |\sqrt{\hat{g}}| \vect{B}\cdot\vect{\Xi}}{|\nabla\Psi|^2} 
\\
& + \mu_0 \frac{dp}{d\Psi} \left[ s_\psi \frac{d^2 V}{d\Psi^2} - \mu_0 \frac{dp}{d\Psi} \iint \frac{d\theta \, d\varphi |\sqrt{\hat{g}}|}{B^2} \right] \iint \frac{d\theta \, d\varphi |\sqrt{\hat{g}}| B^2}{|\nabla\Psi|^2}
\nonumber \\
&+ \left[ \iint \frac{d\theta\, d\varphi |\sqrt{\hat{g}}|\mu_0\vect{J}\cdot\vect{B}}{|\nabla\Psi|^2}\right]^2
-\left[ \iint \frac{d\theta\, d\varphi |\sqrt{\hat{g}}| B^2}{|\nabla\Psi|^2}\right] 
\left[ \iint \frac{d\theta\, d\varphi |\sqrt{\hat{g}}| (\mu_0\vect{J}\cdot\vect{B})^2}{|\nabla\Psi|^2 B^2} \right],
\nonumber
\end{align}
where $\sqrt{\hat{g}}=(\nabla\Psi\cdot\nabla\theta\times\nabla\varphi)^{-1}$, and $\theta$ and $\varphi$ range over $[0,2\pi]$ in the integrals.
This expression is proportional to the stability criterion stated in 
\citep{Bauer1984}
except for the absolute values around $\sqrt{\hat{g}}$ and for the $s_G$ and $s_\psi$ factors that appear here. Eq (\ref{eq:MercierBauerCorrected}) rigorously generalizes this form of the stability criterion in  to be invariant under both parity transformations.

Finally, we note that the Mercier criterion as stated on page 1201 of \cite{Carreras} is parity-invariant and equivalent to (\ref{eq:MercierBauerCorrected}) if one uses the following sign conventions (using notation from that paper): $s \ge 0$, $g = 1/(\nabla s \cdot \nabla \theta \times \nabla\zeta)$ is allowed to have either sign, and $V$ is allowed to have either sign, with $V = \int_0^s ds' \int_0^{2\pi} d\theta \int_0^{2\pi} d\zeta\, g$. This definition of $V$ differs from ours by a factor $s_G s_\psi$.


\bibliographystyle{jpp}

\bibliography{MercierStabilityNearAxis}

\end{document}